\documentclass[10pt,aps,prd,floatfix,notitlepage,showkeys,showpacs,nofootinbib,superscriptaddress]{revtex4-1}

\usepackage{latexsym}
\usepackage{epsfig}
\usepackage{epstopdf,subfigure}
\usepackage{graphicx}
\usepackage{amssymb}
\usepackage{amsmath}
\usepackage{dcolumn,float}
\usepackage{bm}
\usepackage{color}
\usepackage{comment}
\usepackage{ulem,xcolor}

\graphicspath{{pictures/}}

\begin{document}

\title{ Scalar--Tensor  Teleparallel Wormholes by Noether Symmetries}

\author{Sebastian Bahamonde}\email{sebastian.beltran.14@ucl.ac.uk}
\affiliation{Department of Mathematics, University College London,
Gower Street, London WC1E 6BT, UK }
\author{Ugur Camci}\email{ucamci@akdeniz.edu.tr}
\affiliation{Department of Physics, Faculty of Sciences, Akdeniz
University, 07058, Antalya, Turkey}
\author{Salvatore Capozziello}
\email{capozziello@na.infn.it}
\affiliation{Dipartimento di Fisica "E. Pancini", Universit\'a di Napoli
	\textquotedblleft{Federico II}\textquotedblright, I-80126, Napoli, Italy}
\affiliation{INFN Sez. di Napoli, Compl. Univ. di Monte S. Angelo, Edificio G, Via
	Cinthia, I-80126,
	Napoli, Italy}
\affiliation{Gran Sasso Science Institute (INFN), Via F. Crispi 7, I-67100, L' Aquila, Italy}
	\affiliation{Tomsk State Pedagogical University, ul. Kievskaya, 60, 634061 Tomsk, Russia}
\author{Mubasher Jamil}\email{mjamil@sns.nust.edu.pk}
\affiliation{Department of Mathematics, School of Natural
    Sciences (SNS), National University of Sciences and Technology
    (NUST), H-12, Islamabad, Pakistan}

\date{\today}

\begin{abstract}
\textbf{Abstract:} A gravitational  theory of a
scalar field non-minimally coupled with  torsion and  boundary
term is considered with the aim to construct Lorentzian wormholes. Geometrical
parameters including shape  and redshift functions are obtained for these solutions. We adopt
the formalism of Noether Gauge  Symmetry Approach in order to  find  symmetries,
Lie brackets and invariants (conserved quantities).
Furthermore by imposing specific forms of potential function, we
 are able to calculate metric coefficients and discuss their geometrical behavior.
\end{abstract}

\maketitle

\section{Introduction}

The notion of Lorentzian wormholes (WH) arose when Morris and Thorne explored the
possibility of time travels for humans using the principles of
General Relativity (GR) \cite{moris}. Einstein's theory of GR
predicts that the structure and geometry of spacetime in the
presence of matter is not rigid but it is elastic and deformable. The
more compact the object is, the more strong the curvature of space
is, which essentially leads to the idea of black holes. However in
the later case, the fabric of spacetime loses its meaning at the
curvature singularity. If somehow the formation of singularity is
avoided then it would be possible to travel in and out of the
horizon.

The possibility of such a solution to the Einstein field equations was explored 
for the first time  by Flamm \cite{Flamm} soon  after the
formulation  of  GR, but it was later shown that his solution was
unstable. A typical wormhole is a tube-like structure which is
asymptotically flat from both sides. The radius of the wormhole
throat could be  constant or variable depending on its construction
and it  is termed static or non-static respectively.  GR predicts that
to form a WH, an exotic form of matter (violating the energy
conditions) must be present near the throat of the WH. The problem
is the dearth of  reasonable sources  sustaining the wormhole
geometry. One possible candidate is the phantom energy (which is a
cosmic dynamical scalar field with negative kinetic energy in its
Lagrangian) and it is one of the candidates for explaining cosmic
accelerated expansion as well \cite{phantom}. Since the existence of
phantom energy is questionable and no other suitable exotic matter
candidate is available at the moment, an alternative approach is commonly
followed: investigation if the modifications of laws of gravity
(i.e. GR), proposed primarily for explanation of accelerated
expansion and avoiding singularities, can support the WH geometries.
Since the WH is a non-vacuum solution of Einstein field equations,
the presence of some form of energy-matter is necessary to construct
a WH. In the framework of modified gravity, the matter content is
assumed to satisfy the energy conditions near the WH throat, while
 higher curvature correction terms in the Lagrangian are required
to sustain the WH geometry.

Like $f(R)$ (where $R$ is Ricci scalar) gravity which is based on a
symmetric connection, the $f(T)$ (where $T$ is torsion of spacetime)
gravity is based on a skew-symmetric connection \cite{ca}. The latter
theory is one of the many alternative (or modified) theories of
gravity available in literature \cite{c1, geom, rev1,rev2}. $f(T)$ theory is based on
the tetrad formalism and the governing equations are derived by
varying the gravitational Lagrangian with respect to the tetrads.
Since the tetrad is not unique, hence the governing equations
acquire different forms and hence different solutions in different
tetrads (see details in Sec.~II). Like any other theory, $f(T)$
theory have several drawbacks: it does not obey local Lorentz
invariance; violates the first law of thermodynamics; no unique
governing field equations \cite{c2}. Despite these problems, the
theory efficiently describes the cosmic accelerated expansion and
predicts the existence of new kinds of black holes and wormholes. It
can also resolve the dark energy and dark matter conundrums
\cite{cb}. Recently some attempts have been made to unify both
$f(R)$ and $f(T)$ theories as $f(R,T)$  gravity o by introducing a boundary term $B$ in $f(T,B)$ gravity \cite{Bahamonde:2015zma,Bahamonde:2016kba,c3}. 

Here,  we want to study wormholes in a theory
where torsion is non-minimally coupled with a scalar field  and includes a matter action. It is important to note
that similar models have also been studied in literature \cite{cc}.

In this paper, we employ the Noether Symmetry Approach \cite{cimall} and a
wormhole metric ansatz to obtain the governing system of
differential equations. After solving the equations, we get the form
of metric coefficients, symmetry generators, invariants and the form
of torsion and potential functions. This approach has been
extensively studied in literature \cite{ns,and1,and2,Kucukakca:2014vja,Kucukakca:2013mya}. 

The plan of the paper
is as follows: In Sec. II, we provide a brief review 
of generalized teleparallel gravity and sketch relevant notations.
In Sec.~III, we present the model starting with the action of a
non-minimally coupled scalar field with both torsion and the
boundary term. We also derive the field equations and choose an
ansatz for the wormhole metric. In Sec. IV and its sub-sections, we
construct a point-like Lagrangian, search for Noether symmetries and derive the 
wormhole solutions. Discussion and conclusions are given in Sec. V.

\section{Teleparallel equivalent of general relativity}
\label{tele}

In what follows we will follow  conventions outlined
in~\cite{jimi}. Here we present a brief review of the
teleparallel formalism of GR. Unlike GR, the dynamical variable in the teleparallel theory is the
tetrad $e^{a}_{\mu}$ (rather than the metric tensor), where Greek
and Latin indices denote spacetime and tangent space indices
respectively. Here the metric tensor $g_{\mu\nu}$ is related with
the tetrads as
\begin{align}
g_{\mu\nu}=e^{a}_{\mu}e^{b}_{\nu}\eta_{ab},
\end{align}
where $\eta_{ab}$ denotes the Minkowski metric tensor. The inverse
tetrad $E^{\mu}_{a}$ is defined  as
\begin{align}
E_{m}^{\mu}e_{\mu}^{n}=\delta^{n}_{m}, \quad {\text {and}} \quad
E_{m}^{\nu}e_{\mu}^{m}=\delta^{\nu}_{\mu}.\label{deltamunu}
\end{align}
Here $e$ is  the determinant of the tetrad $e^a_\mu$,
which can be evaluated from the determinant of the metric tensor
$e=\sqrt{-g}$.

GR is  based on the symmetric Levi-Civita
connection, whereas, teleparallel gravity relies on the
anti-symmetric Weitzenb\"ock connection $W_{\mu}\,^{a}\,_{\nu}$
defined as
\begin{align}
W_{\mu}{}^{a}{}_{\nu}=\partial_{\mu}e^{a}{}_{\nu}.
\end{align}
The anti-symmetric nature of  connection leads to the concept of
torsion in geometry. The torsion tensor is  the antisymmetric
part of the Weitzenb\"ock connection
\begin{align}
T^{a}\,_{\mu\nu}&=W_{\mu}{}^{a}{}_{\nu}-W_{\nu}{}^{a}{}_{\mu}=\partial_{\mu}e_{\nu}^{a}-\partial_{\nu}e_{\mu}^{a}.
\end{align}
The tensor $T_{\mu}$, referred as the torsion vector, is defined through the  contraction  of the torsion tensor, i.e.
\begin{align}
T_{\mu}=T^{\lambda}\,_{\lambda\mu}.
\end{align}
To construct the field equations of teleparallel gravity, we
consider the following Lagrangian density and vary it with respect
to the tetrad
\begin{align}
\mathcal{L}_T= \frac{e}{2\kappa^2}S^{abc}T_{abc}, \label{taction}
\end{align}
where
\begin{align}
S^{abc}=\frac{1}{4}(T^{abc}-T^{bac}-T^{cab})+\frac{1}{2}(\eta^{ac}T^b-\eta^{ab}T^c).
\end{align}
The torsion scalar $T$ is defined as
\begin{align}
T=S^{abc}T_{abc}.
\end{align}

Furthermore, to express teleparallel gravity as an equivalent of GR, we write the Levi-Civita connection ${}^0
\Gamma$ in terms of the Weitzenb\"ock connection as
\begin{align}
{}^0 \Gamma^{\mu}_{\lambda\rho}=W_{\lambda}{}^{\mu}{}_{\rho}-K_{\lambda}{}^{\mu}{}_{\rho}, \label{Levi}
\end{align}
where here $K$ is called the contortion tensor and it is defined as
\begin{align}
2K_{\mu}\,^{\lambda}\,_{\nu}&=T^{\lambda}\,_{\mu\nu}-T_{\nu\mu}\,^{\lambda}+T_{\mu}\,^{\lambda}\,_{\nu}.
\end{align}
This contortion tensor is antisymmetric in its last two indices. Now
expressing the Ricci scalar of the Levi-Civita connection in terms
of the Weitzenb\"ock connection; using~(\ref{Levi}), we get
\begin{align}
  R = - T + \frac{2}{e}\partial_\mu (e T^\mu) \,. \label{ricciT}
\end{align}
As the difference between the Ricci scalar and the torsion scalar is
simply a total derivative, the action~(\ref{taction}) gives rise to
the same dynamics as the Einstein Hilbert action. This shows that
teleparallel gravity is indeed equivalent to GR.
Defining the boundary quantity
\begin{align}
B=\frac{2}{e}\partial_\mu (e T^\mu)
\end{align}
one then has simply the relation $R=-T+B$. Note that one can write
$B$ in terms of a Levi-Civita covariant derivative simply as
$B=2\nabla_\mu T^\mu$.

\section{Non-minimally coupled scalar field to  torsion and  boundary term}
Let us consider the following gravitational action which describes a non-minimally coupled scalar field to both torsion and the boundary term \cite{Bahamonde:2015hza,Zubair:2016uhx},
\begin{align}
S = \int
  \left[
    \frac{1}{2}( f(\phi) T +g(\phi) B+\partial_\mu \phi \partial^\mu \phi )-V(\phi) %+ L_{\rm m}
  \right] e\, d^4x.\label{action}
\end{align}
Here, $V(\phi)$ is the scalar field potential  and $f(\phi)$ and $g(\phi)$ are smooth functions of the scalar field $\phi$.
This action is not a new theory, instead it is a generalisation or rather a
	unification of different theories into one action. This theory
	is very rich in the sense that one can recover very well-known
	scalar-tensor theories.  In fact, let us for example choose 
	\begin{eqnarray}
	f(\phi)=1-\xi \phi^2\,, \ \ g(\phi)=-\chi  \phi^2\,,\label{1}
	\end{eqnarray}
	where $\xi$ and $\chi$ are coupling constants. Depending on the value of $\chi$ and $\xi$, we can recover scalar tensor theories non-minimally coupled with the torsion scalar ($\chi=0$), with the boundary term ($\xi=0$), with the Ricci scalar ($\chi=-\xi$) and also quintessence theory ($\xi=\chi=0$).	The latter theories have been widely studied in the literature from cosmology to astrophysical sources like wormholes. Traversable wormholes supported by non-minimally coupled scalar field with the Ricci scalar ($\chi=-\xi$) were first studied in Ref. \cite{Barcelo:2000zf}. They found that depending on the value of the coupling constant, one can construct different kind of wormholes. Wormholes supported by a kink-like configuration of a scalar field was studied in  \cite{Sushkov:2002ef}. Moreover, in \cite{Bronnikov:2002sf,Bronnikov:2005an}, the authors 
	studied the stability of electrically charged and neutral wormholes within this theory.  \\
Additionally, canonical Brans-Dicke scalar tensory theory with $w_{BD}=1$ can be recovered if we set
	\begin{eqnarray}\label{2}
	f(\phi)=\xi \phi^2\,, \ \ g(\phi)=-\chi \phi^2\,
	\end{eqnarray}
with $\chi=-\xi$. Wormholes configurations under this theory have been widely studied in the literature \cite{brans}. One interesting feature of  Bans-Dicke wormholes is that they can satisfy the energy conditions. For the reader interested on cosmology on all these theories mentioned above, see Refs~\cite{Torres:2002pe,Bahamonde:2015hza,Boisseau:2000pr}. In the context of teleparallel gravity, wormhole solutions have also been studied (see \cite{Bohmer:2011si} and their referees therein), but to the best of our knowledge, wormholes in teleparallel scalar-tensor theories (for example with $\xi\neq -\chi$) have not been considered yet until now. Note that action is very rich since we can study all these kind of theories and then make the corresponding choice of the coupling functions. Moreover, using this approach, we can have a direct relation from the teleparallel and metric counterparts.\\
By varying this action with respect to the tetrad field we find the following field equations
\begin{align}
2f(\phi)\left[ e^{-1}\partial_\mu (e S_{a}{}^{\mu\nu})-E_{a}^{\lambda}T^{\rho}{}_{\mu\lambda}S_{\rho}{}^{\nu\mu}-\frac{1}{4}E^{\nu}_{a}T\right]-E^{\nu}_a \left[\frac{1}{2}\partial_\mu \phi \partial^\mu \phi -V(\phi)\right]
\nonumber\\ +E^{\mu}_a \partial^\nu \phi \partial_\mu \phi + 2(\partial_{\mu}f(\phi)+\partial_{\mu}g(\phi)) E^\rho_a S_{\rho}{}^{\mu\nu}+E^{\nu}_{a}\Box g(\phi)-E^\mu_a \nabla^{\nu}\nabla_{\mu}g(\phi)=0\,. %T^\nu_a, \label{fieldeqn}
\end{align}
%Here, $\mathcal{T}_{a}^{\nu}=E_{a}^{\lambda}\mathcal{T}_{\lambda}^{\nu}$ is the energy-momentum tensor.
By contracting this equation with the tetrad field $e_{\lambda}^{a}$, we can have this equation only in space-time index
\begin{align}
2f(\phi)\left[ e^{-1}e_{\lambda}^{a}\partial_\mu (e S_{a}{}^{\mu\nu})-T^{\rho}{}_{\mu\lambda}S_{\rho}{}^{\nu\mu}-\frac{1}{4}\delta^{\nu}_{\lambda}T\right]-\delta^{\nu}_{\lambda} \left[\frac{1}{2}\partial_\mu \phi \partial^\mu \phi -V(\phi)\right]
\nonumber\\ +\partial^\nu \phi \partial_\lambda \phi + 2(\partial_{\mu} f(\phi)+\partial_{\mu}g(\phi)) S_{\lambda}{}^{\mu\nu}+\delta^{\nu}_{\lambda}\Box g(\phi)- \nabla^{\nu}\nabla_{\lambda}g(\phi)=0\,. %\mathcal{T}^\nu_\lambda. \label{fieldeqn2}
    \end{align}
Now, by varying the action (\ref{action}) with respect to the scalar field we obtain the so-called, Klein-Gordon equation which reads
\begin{eqnarray}
\Box \phi+V'(\phi)&=\frac{1}{2}\Big(f'(\phi)T+g'(\phi)B\Big).\label{KG}
\end{eqnarray}
Let us consider a static spherically symmetric space-time in which the metric is
\begin{align}
ds^2=e^{a(r)}dt^2-e^{b(r)} dr^{2}-M(r)^{2}d\Omega^{2}, \label{metric}
\end{align}
where $a(r), b(r)$ and $M(r)$ are functions of the radial coordinate $r$, and $d\Omega^2 = d\theta^2 + \sin^2 \theta d\varphi^2$. \\
%The corresponding diagonal tetrad will be
%\begin{align}
%e_{\mu}^{a}&=\textrm{diag}\Big(e^{a(r)/2},-e^{b(r)/2},-r,-r\sin\theta\Big).\label{tetrad1}
%\end{align}
The corresponding off-diagonal tetrad field to the metric (\ref{metric}) is
\begin{eqnarray}
e^{a}_{\mu}= \left(\begin{array}{cccc}
e^{a/2}&0&0&0\\
0&e^{b/2}\sin\theta\cos\varphi & M(r) \cos\theta\cos\varphi & -M(r) \sin\theta\sin\varphi\\
0&e^{b/2}\sin\theta\sin\varphi & M(r) \cos\theta\sin\varphi & M(r) \sin\theta\cos\varphi\\
0& e^{b/2}\cos\theta & -M(r)\sin\theta & 0
\end{array}\right)\label{tetrad2}\; .
\end{eqnarray}
This tetrad field has been used in the context of $f(T)$ gravity since it does not produce a constraint, namely $f_{TT}=0$, as a diagonal tetrad.
%Using the tetrad (\ref{tetrad1}), the field equations become
%\begin{eqnarray}
%V(\phi)+\frac{1}{2} e^{-b(r)} \phi '^2-\frac{e^{-b(r)} }{2 r}\left(r b'(r) g'(\phi)+4 f'\right)+\frac{ f(\phi)}{r}\Big(e^{-b(r)} \left(b'(r)-\frac{1}{r}\right)+\frac{1}{r}\Big)+e^{-b(r)} g''(\phi)&=&0\label{eq1}\,\\
%V(\phi)-\frac{1}{2} e^{-b(r)} \phi '^2+\frac{f(\phi)}{r}\Big(\frac{1}{r}-e^{-b(r)} \left(a'(r)+\frac{1}{r}\right)\Big)+e^{-b(r)}\left(\frac{a'(r)}{2}+\frac{2}{r}\right) g'(\phi )\,&=&0\label{eq2},\\
%V(\phi )+\frac{1}{2} e^{-b(r)} \phi '^2-e^{-b(r)}\left(\frac{a'(r)}{2}+\frac{1}{r}\right) f'(\phi)+e^{-b(r)}\Big(g''(\phi)-\frac{1}{2} b'(r)g'(\phi)\Big)&&\nonumber\\
%-\frac{1}{2} e^{-b(r)} f(\phi)\Big(a''(r)+\frac{1}{2} a'(r) \left(a'(r)-b'(r)\right)+\frac{a'(r)}{r}-\frac{b'(r)}{r}\Big)&=&0\label{eq3}\,.
%\end{eqnarray}
For this tetrad, the field equations read
\begin{eqnarray}
V(\phi)+\frac{1}{2} e^{-b(r)} \phi '^2+\frac{e^{-b(r)} }{2 r}\left(r b'(r) g'(\phi)+4 f'\right)-\frac{ f(\phi)}{r}\Big(e^{-b(r)} \left(b'(r)-\frac{1}{r}\right)+\frac{1}{r}\Big)-e^{-b(r)} g''(\phi)&&\nonumber\\
-\frac{2 e^{-\frac{b(r)}{2}} }{r}\phi' \left(f'(\phi)+g'(\phi)\right)&=&0, \label{eq1b}\\
V(\phi)-\frac{1}{2} e^{-b(r)} \phi '^2-\frac{f(\phi)}{r}\Big(\frac{1}{r}-e^{-b(r)} \left(a'(r)+\frac{1}{r}\right)\Big)-e^{-b(r)}\left(\frac{a'(r)}{2}+\frac{2}{r}\right) g'(\phi )\,&=&0, \label{eq2b}\\
V(\phi )+\frac{1}{2} e^{-b(r)} \phi '^2+e^{-b(r)}\left(\frac{a'(r)}{2}+\frac{1}{r}\right) f'(\phi)-e^{-b(r)}\Big(g''(\phi)-\frac{1}{2} b'(r)g'(\phi)\Big)&& \nonumber\\
+\frac{1}{2} e^{-b(r)} f(\phi)\Big(a''(r)+\frac{1}{2} a'(r) \left(a'(r)-b'(r)\right)+\frac{a'(r)}{r}-\frac{b'(r)}{r}\Big)-\frac{ e^{-\frac{b(r)}{2}} }{r}\phi' \left(f'(\phi)+g'(\phi)\right)&=&0\,.\label{eq3b}
\end{eqnarray}
Here, primes denotes differentiation with respect to the radial coordinate $r$. It can be shown that if we choose $f(\phi)=1+\xi \phi^2=-g(\phi)$ we recover the case studied in \cite{Barcelo:2000zf,Sushkov:2002ef}. \\
%\textcolor{red}{PROBLEM HERE: IN THE CASE STUDIED IN \cite{Sushkov:2002ef}, the addional term is zero, so I can not know if the equations have this term or not. I think that both equations (\ref{eq1})-(\ref{eq3}) and (\ref{eq1b})-(\ref{eq3b}) are correct. They are different because as in f(T) gravity, different tetrads give different equations (that's why one needs to be very careful when we use tetrads \cite{Tamanini:2012hg})}\\
Additionally, the Klein-Gordon equation becomes
\begin{align}
-e^{-b(r)} \phi '(r) \left(\frac{1}{2} a'(r)-\frac{b'(r)}{2}+\frac{2}{r}\right)-e^{-b(r)} \phi'' + V'(\phi)-\frac{1}{2} \left(B(r) g'(\phi )+T(r) f'(\phi)\right)&=0, \label{KG2}
\end{align}
%where the torsion scalar and the boundary term are given by (\textcolor{red}{Using the tetrad (\ref{tetrad1})})
%\begin{align}
%T(r)&=\frac{2 e^{-b(r)} }{r^2}\left(r a'(r)+1\right)\,\label{T1}\\
%B(r)&=-\frac{2}{r^2}+e^{-b(r)} \left(a''(r)-\frac{1}{2} a'(r) b'(r)+\frac{1}{2} a'(r)^2+\frac{4 a'(r)}{r}-\frac{2 b'(r)}{r}+\frac{4}{r^2}\right)\,.\label{B1}
%\end{align}
where the torsion scalar and the boundary term are respectively
given by
\begin{align}
T(r)&=\frac{2 e^{-b(r)}}{M^2}\left(e^{\frac{b(r)}{2}}- M'\right) \left(-M a' - M' + e^{\frac{b(r)}{2}} \right)\,,\label{T2}\\
B(r)&= -2 e^{-\frac{b(r)}{2}} \left(\frac{a'}{M} + \frac{2 M'}{M^2}\right) + e^{-b(r)} \left[ a''-\frac{1}{2} a' b' + \frac{1}{2} a'^2 + \frac{4 M' a'}{M}-\frac{2 M' b'}{M} + \frac{4 M'^2}{M^2} + \frac{4 M''}{M} \right]\,.\label{B2}
\end{align}
%\textcolor{red}{If we compute $-T(r)+B(r)$ using (\ref{T1}) and (\ref{B1}) or (\ref{T2}) and (\ref{B2}) we find the same Ricci scalar (the correct result):}
Clearly, by substracting (\ref{B2}) with (\ref{T2}) we recover the Ricci scalar
\begin{align}
R(r)&=-T(r)+B(r)=-\frac{2}{M^2}+e^{-b(r)} \left[ a'' - \frac{1}{2} a' b' + \frac{1}{2} a'^2 + \frac{2 M' a'}{M}-\frac{2 M' b'}{M} + \frac{2 M'^2}{M^2} + \frac{4 M''}{M} \right]\,.
\end{align}

\section{Wormhole solutions}
Now, we have all the ingredients to study wormhole configurations in this framework. A spherically symmetric wormhole will be given by choosing
\begin{align}
e^{b(r)}&=\Big(1-\frac{\beta (r)}{r}\Big)^{-1}\,, \label{worm-b}
\end{align}
where $\beta(r)$ is the shape function of the wormhole and $a(r)$ is known as
the redshift function. Thus, the field equations read
\begin{eqnarray}
V(\phi)+\left(1-\frac{\beta (r)}{r}\right)\Big(\frac{1}{2} \phi '^2+\frac{2}{r}f'(\phi)-g''(\phi)\Big)-\frac{ \beta '(r)}{r^2}f(\phi)+\left(\beta '(r)-\frac{\beta (r)}{r}\right) \frac{g'(\phi)}{2 r} & &  \nonumber\\
-\frac{2  \left(f'(\phi)+g'(\phi )\right)}{r} \sqrt{1-\frac{\beta(r)}{r}}= 0, \label{33} && \quad  \\
V(\phi)-\frac{1}{2} \phi '^2-\left(\beta (r) a'(r)-r a'(r)+\frac{\beta (r)}{r}\right)\frac{f(\phi) }{r^2}-\left(\frac{a'(r)}{2}-\frac{\beta (r) a'(r)}{2 r}-\frac{2 \beta (r)}{r^2}+\frac{2}{r}\right)g'(\phi)+\frac{\beta (r) }{2 r}\phi '(r)^2=0, \label{34} && \quad  \\
V(\phi)+\left(1-\frac{\beta (r)}{r}\right) \left(\frac{1}{2} a''(r) f(\phi)+\frac{1}{2} a'(r) f'(\phi)+\frac{f'(\phi)}{r}+\frac{1}{2} \phi '^2\right)-g''(\phi)+\frac{\left(r a'(r)+2\right)}{4 r} a'(r) f(\phi)&&\nonumber\\
-\left(\frac{1}{2} a'(r)^2+\frac{a'(r)}{2 r}-\frac{1}{r^2}\right)\frac{ \beta (r) f(\phi)}{2 r}-\left(\frac{a'(r)}{2}+\frac{1}{r}\right)\frac{ f(\phi) \beta '(r)}{2 r}+\frac{\beta '(r) g'(\phi)}{2 r}&&\nonumber\\
- \left(\frac{g'(\phi)}{2 r}-g''(\phi)\right)\frac{\beta (r)}{r}-\frac{ \left(f'(\phi)+g'(\phi )\right)}{r} \sqrt{1-\frac{\beta(r)}{r}}=0 \label{35}. & & \quad
\end{eqnarray}
In the following section, we will use the Noether symmetry approach to find analytical wormhole solutions within this theory.
\subsection{The point-like Lagrangian}
The action (\ref{action}) can be written as $S = \int \mathcal{L} \,dr$, where $\mathcal{L}$ is the Lagrangian density. In the background of the static spherically symmetric space-time (\ref{metric}), the point-like Lagrangian density for the teleparallel theory of gravity is obtained as
\begin{eqnarray}
& & \mathcal{L} = f(\phi) e^{\frac{a-b}{2}} \left( e^{b/2} - M' \right) \left[  e^{b/2} -M a' - M' \right] - g(\phi) e^{a/2} \left(M a' + 2 M' \right) \nonumber \\& & \qquad \quad - g_{,\phi}  e^{\frac{a-b}{2}} \left( \frac{M^2}{2} a' \phi' + 2 M M' \phi' \right) - \frac{M^2}{2} e^{\frac{a-b}{2}} \phi'^2 - M^2 e^{\frac{a+b}{2}} V(\phi) %+ M^2 e^{(a+b)/2} \mathcal{L}_m
\,, \label{lagr1}
\end{eqnarray}
where $g_{,\phi} = d g(\phi) / d\phi$. Note that the hessian determinant of the above Lagrangian is zero. This is clearly due to the absence of the generalized velocity $b'$ in the point-like Lagrangian. By varying the point-like Lagrangian density (\ref{lagr1}) with respect to the redshift function $a$, we get
\begin{eqnarray}
& & f(\phi)  \left[ \frac{2 M''}{M} + \frac{M'^2}{M^2} -\frac{b' M'}{M} - \frac{e^b}{M^2} \right]  + 2 f_{,\phi} \frac{\phi'}{M} \left( M' - e^{b/2} \right)  -  g_{,\phi} \left( \phi'' + 2 e^{b/2} \frac{\phi'}{M} \right) + \frac{1}{2} \phi'^2 + e^{b} V_{,\phi} = 0\,. \label{eq11}
\end{eqnarray}
Variations with respect to the shape function $b$ gives us
\begin{eqnarray}
& &  f (\phi) \left[ \frac{M'}{M} \left( a' + \frac{M'}{M} \right) - \frac{e^b}{M^2} \right] -  g_{,\phi} \left( \frac{a' \phi'}{2} + \frac{ 2 M' \phi'}{M} \right) - \frac{1}{2} \phi'^2 + e^{b} V (\phi) = 0\,. \label{eq22}
\end{eqnarray}
Now, if we vary the point-like Lagrangian density with respect to $M$, we find
\begin{eqnarray}
& & f(\phi)  \left[ \frac{M''}{M} + \frac{a''}{2} + (a' - b') \left( \frac{a'}{4} + \frac{M'}{2 M} \right) \right]  +  f_{,\phi} \phi' \left[ \frac{a'}{2} + \frac{1}{M} \left( M' - e^{b/2} \right) \right] \nonumber \\& & \quad +  g_{,\phi} \left[ \phi' \left( \frac{b'}{2} - \frac{ e^{b/2}}{M} \right) - \phi'' \right] + \left( \frac{1}{2}- g_{,\phi \phi} \right) \phi'^2  + e^{b} V_{,\phi} = 0\,. \label{eq33}
\end{eqnarray}
Finally, by varying the point-like Lagrangian density with respect to $\phi$ yields
\begin{eqnarray}
& & \frac{e^b}{2} \left[ f_{,\phi} T(r) + g_{,\phi} B(r) \right] + \phi'' + \phi' \left[ \frac{1}{2} (a' - b') + \frac{2 M'}{M} \right] - e^{b} V_{,\phi} = 0, \label{eq44}
\end{eqnarray}
The latter equation is the Klein-Gordon equation.\\
The {\it energy function} associated with $\mathcal{L}$ is defined by
\begin{equation}
E_{\mathcal{L}} = q'^k \frac{\partial \mathcal{L}}{\partial q'^k} - \mathcal{L}, \label{energy}
\end{equation}
which is also the {\it Hamiltonian} of the system. Here, $q^i, \, i=1,2,3,4$, are the generalized coordinates where $q^i = \{ a, b, M, \phi \}$, for the Lagrangian density (\ref{lagr1}) of teleparallel theory of gravity. Then $E_{\mathcal{L}}$ has the following form
\begin{equation}
E_{\mathcal{L}} = M^2 e^{(a-b)/2} \left[ f(\phi)  \left( \frac{M'^2}{M^2} + \frac{a' M'}{M} -\frac{e^b}{M^2} \right) - g_{,\phi} \left( \frac{1}{2} a' \phi' + \frac{2 M' \phi'}{M} \right) - \frac{1}{2} \phi'^2 + e^b V(\phi)  \right], \label{energy2}
\end{equation}
which vanishes because of  Eq.~(\ref{eq22}) due to the variation with respect to $b$, i.e. $E_{\mathcal{L}} = 0$. This can be explicitly solved in terms of $b$ as a function of the remaining generalized coordinates:
\begin{equation}
e^b = \frac{M^2 \left[ - f(\phi)  \left( \frac{M'^2}{M^2} + \frac{a' M'}{M} \right) + g_{,\phi} \left( \frac{1}{2} a' \phi' + \frac{2 M' \phi'}{M} \right) + \frac{1}{2} \phi'^2 \right]}{M^2 V(\phi) - f(\phi) }. \label{coord-b}
\end{equation}
Then, using (\ref{worm-b}), the shape function $\beta(r)$ of the wormhole takes the form
\begin{equation}
\beta(r) = \frac{r \left[ - f(\phi)  \left( \frac{M'^2}{M^2} + \frac{a' M'}{M} - \frac{1}{M^2}\right) + g_{,\phi} \left( \frac{1}{2} a' \phi' + \frac{2 M' \phi'}{M} \right) + \frac{1}{2} \phi'^2 + V(\phi) \right]}{\left[ - f(\phi)  \left( \frac{M'^2}{M^2} + \frac{a' M'}{M} \right) + g_{,\phi} \left( \frac{1}{2} a' \phi' + \frac{2 M' \phi'}{M} \right) + \frac{1}{2} \phi'^2 \right] }. \label{shape-1}
\end{equation}
In its present form, $\beta$ is expressed in terms of several
arbitrary functions. Below, we shall determine explicitly $\beta$
 as a function of radial coordinate and  qualitatively investigate its
behavior.

\subsection{The Noether Symmetry Approach}
Let us consider a Noether symmetry vector generator  \cite{cimall} 
\begin{equation}
{\bf X} = \xi \frac{\partial}{\partial r} + \eta^i \frac{\partial}{\partial q^i}, \label{ngs-gen}
\end{equation}
where $q^i = \{ a, b, M, \phi \}$ are the generalized coordinates in the configuration space ${\cal Q }\equiv \{ q^i , i=1,\ldots, 4 \}$ of the Lagrangian, whose tangent space is ${\cal TQ }\equiv \{q^i,q'^i\}$. The components $\xi$ and $\eta^i$ of the Noether symmetry generator ${\bf X}$ are functions of $r$ and $q^i$. The existence of a Noether symmetry implies the existence of a vector field ${\bf X}$ given in (\ref{ngs-gen}) if the Lagrangian $ \mathcal{L}(r, a, b, M, \phi, a', b', M', \phi' )$ satisfies
\begin{equation}
{\bf X}^{[1]} \mathcal{L} + \mathcal{L} ( D_r \xi) = D_r G\,, \label{ngs-eq}
\end{equation}
where ${\bf X}^{[1]}$ is the first prolongation of the generator (\ref{ngs-gen}) in such a form
\begin{equation}
{\bf X}^{[1]} = {\bf X}  + \eta'^i \frac{\partial}{\partial q'^i},
\end{equation}
$G(r, q^i)$ is a gauge function, $D_r$ is the total derivative operator with respect to $r$, $
D_r =\partial / \partial r + q'^i \partial / \partial q^i$, and $\eta'^i$ is defined as $\eta'^i = D_r \eta^i - q'^i D_r \xi$. It is important to give the following Noether first integral to emphasize the significance of Noether symmetry that if ${\bf X}$ is the Noether symmetry generator corresponding to the Lagrangian $\mathcal{L}(r,q^i,q'^i)$, then
\begin{equation}
I =- \xi E_{\mathcal{L}} + \eta^i \frac{\partial \mathcal{L}}{\partial q'^i} - G\,,
\end{equation}
which is also the Hamiltonian or a conserved quantity associated with the generator ${\bf X}$. Now we seek for the condition in order that the Lagrangian density (\ref{lagr1}) would admit any Noether symmetry. The Noether symmetry condition (\ref{ngs-eq}) for the Lagrangian (\ref{lagr1}) gives rise to the following set of differential equations
\begin{eqnarray}
& & \xi_{,a} = 0, \quad \xi_{,b} = 0, \quad \xi_{,M} = 0, \quad \xi_{,\phi} = 0, \nonumber \\& & 2 f \eta^3_{,a} - g_{,\phi} M \eta^4_{,a} = 0, \quad 2 f \eta^3_{,b} - g_{,\phi} M \eta^4_{,b} = 0, \nonumber \\& & \frac{1}{2} M (\eta^1 - \eta^2 ) + 2 \eta^3 + g_{,\phi} \left( M \eta^1_{,\phi} + 4 \eta^3_{,\phi} \right) + M ( 2 \eta^4_{,\phi} - \xi_{,r} ) = 0, \nonumber \\& & f \left( \frac{1}{2} M \eta^1_{,b} + \eta^3_{,b} \right) - g_{,\phi} M \eta^4_{,b} = 0, \quad g_{,\phi} \left( \frac{1}{2} M \eta^1_{,b} + 2 \eta^3_{,b} \right) + M \eta^4_{,b} = 0, \nonumber \\& & f \left[ \frac{1}{2} ( \eta^1 -\eta^2) + M \eta^1_{,M} + 2 \eta^3_{,M} - \xi_{,r} \ \right] + f_{,\phi} \eta^4 - 2 g_{,\phi} M \eta^4_{,M} = 0, \nonumber \\& & g_{,\phi} \left( M \eta^4_{,a} + \frac{1}{4} M^2 \eta^4_{,M} \right) - f \left[ \frac{1}{4} M (\eta^1 - \eta^2) + \frac{1}{2} \eta^3 + \eta^3_{,a} + \frac{1}{2} M (\eta^1_{,a} + \eta^3_{,M}) \right] - \frac{1}{2} f_{,\phi} \eta^4 = 0, \nonumber  \\& & g_{,\phi} \left[ \frac{1}{2} M (\eta^1 - \eta^2 ) + 2 \eta^3 + 4 \eta^3_{,a} + M ( \eta^1_{,a} + \eta^4_{,\phi} - \xi_{,r} ) \right] + g_{,\phi \phi} M \eta^4 + 2 M\eta^4_{,a} - 2 f \eta^3_{,\phi} = 0,  \label{ngs-eqs} \\& & g_{,\phi} \left[ \frac{1}{2} M (\eta^1 - \eta^2 ) + \eta^3 + \frac{1}{4} M^2  \eta^1_{,M} + M ( \eta^3_{,M} + \eta^4_{,\phi} - \xi_{,r} ) \right] + g_{,\phi \phi} M \eta^4 - \frac{1}{2} f M \eta^1_{,\phi}-  f \eta^3_{,\phi} + \frac{1}{2} M^2 \eta^4_{,M} = 0, \nonumber\\& & (f + g) \left[ \frac{1}{2} M \eta^1 + \eta^3 + M \eta^1_{,a} + 2 \eta^3_{,a} \right] + M (f+g)_{,\phi} \eta^4 - f M e^{-b/2} \eta^3_{,r} + \frac{1}{2} g_{,\phi} M^2 e^{-b/2} \eta^4_{,r} + e^{-a/2} G_{,a} = 0, \nonumber\\& & (f + g) \left[ M \eta^1_{,b} + 2 \eta^3_{,b} \right] + e^{-a/2} G_{,b} = 0, \nonumber \\& & (f + g) \left( \eta^1 +  M \eta^1_{,M} + 2 \eta^3_{,M} \right) + 2 (f+g)_{,\phi} \eta^4 - f M e^{-b/2} \eta^1_{,r} - f e^{-b/2} \eta^3_{,r} + 2 g_{,\phi} M e^{-b/2} \eta^4_{,r} + e^{-a/2} G_{,M} = 0, \nonumber\\& & (f + g) \left[ M \eta^1_{,\phi} + 2 \eta^3_{,\phi} \right] + g_{\phi} e^{-b/2} \left( \frac{1}{2} M^2 \eta^1_{,r} + 2 M \eta^3_{,r} \right) + M^2 e^{-b/2} \eta^4_{,r} + e^{-a/2} G_{,\phi} = 0, \nonumber \\& & (M^2 V(\phi) - f) \left[ \frac{1}{2} (\eta^1 + \eta^2) - \xi_{,r} \right] + 2 M V(\phi) \eta^3 +  (M^2 V(\phi)_{,\phi} - f_{,\phi}) \eta^4  \nonumber \\& & \qquad + e^{-b/2}(f + g) ( M \eta^1_{,r} + 2 \eta^3_{,r} ) + e^{-(a+b)/2} G_{,r} = 0. \nonumber
\end{eqnarray}
It is explicitly seen from the above equations that $\xi = \xi(r)$, and a trivial solution is $\xi = const., \eta^i = 0$ and $G = const$. This trivial solution means that ${\bf X}_1 = \partial_r$ is a Noether symmetry for any form of the $f(\phi), g(\phi)$ and $V(\phi)$. The corresponding Noether constant of ${\bf X}_1$ is $I = - E_{\mathcal{L}}$ which vanishes due to the Eq. (\ref{eq22}).
%In the following we consider non-trivial solutions of the Noether symmetry conditions.

{\bf Case (i):} $V(\phi) = 0$ (vanishing potential).

Relevant subcases:

{\bf (i.a)}. Take $f(\phi) = f_0 - \zeta \phi^2$ and $g(\phi)= -\chi \phi^2$, where $f_0, \zeta$ and $\chi$ are constants.

(i.a.1) If $\chi = 0$, then we have Teleparallel dark energy (non-minimally coupled to $T$).

(i.a.2) If $\chi = - \zeta$, then it means that non-minimally coupled to the Ricci scalar.

(i.a.3) If $\zeta = 0$, then it gives non-minimally coupled to the Boundary term.

(i.a.4) The condition $\zeta = \chi = 0$ gives rise to  \textit{quintessence models}.

{\bf (i.b)}. If $f(\phi) = f_0 - \zeta \phi$ and $g(\phi)= -\chi \phi$, then the possibilities are

(i.b.1) $\chi = 0$, (i.b.2) $\chi = - \zeta$ (Brans-Dicke theory with $w(\phi)=1$), (i.b.3) $\zeta = 0$, and (i.b.4) The condition $\zeta = \chi = 0$ gives rise to \textit{quintessence models} and same as subcase (i.a.4).

\bigskip

The Noether symmetries for the Lagrangian density (\ref{lagr1}) are obtained as
\begin{eqnarray}
& & \xi = c_1, \quad \eta^1 = - 2 c_2, \quad \eta^2 = 2 c_2, \quad \eta^3 = c_2 M, \quad \eta^4 = 0, \quad G = c_3, \label{ngsc-1}
\end{eqnarray}
for the subcases (i.a.1), (i.a.2), (i.a.3), (i.b.1) and (i.b.2) which yields that
\begin{eqnarray} \label{ngsv-1}
& & {\bf X}_1 =  \partial_r, \qquad {\bf X}_2 = - \partial_a + \partial_b + \frac{1}{2} M \partial_M.
\end{eqnarray}
The corresponding first integrals of the above Noether symmetries are
\begin{eqnarray}
& & I_1 = -  E_{\mathcal{L}}, \qquad I_2 =  \frac{1}{2} f(\phi) M^2 e^{(a-b)/2} a', \label{fint-1}
\end{eqnarray}
where $I_1$ vanishes due to the $E_{\mathcal{L}} = 0$. The second of the above first integrals gives $(e^{a/2})' = \alpha e ^{b/2} / \left[ f(\phi) M^2 \right]$, where $\alpha = I_2$. Thus the Eq. $E_{\mathcal{L}} = 0$ takes the form
\begin{eqnarray}
& & u'^2 = \frac{\alpha^2}{(f_0 - \zeta \phi^2)^3 M^2} \left[ 2 P  \frac{u'}{u}  + Q \right],  \label{ode-1}
\end{eqnarray}
where $u$ is defined as $u \equiv e^{a/2}$, $P$ and $Q$ are defined by
\begin{eqnarray}
& & P= \chi \phi \phi' +(f_0 - \zeta \phi^2) \frac{M'}{M}, \quad Q = \frac{M'}{M} \left( P + 3 \chi \phi \phi' \right)- \frac{\phi'^2}{2}.
\end{eqnarray}
The Eq. (\ref{ode-1}) is a non-linear differential equation for $u$, and it could be solved under some assumptions. %If $P \equiv 0$, then

In subcase (i.b.3), i.e. $f(\phi) = f_0$ and $g(\phi) = - \chi \phi$, we have three Noether symmetries ${\bf X}_1, {\bf X}_2$  and ${\bf X}_3 = \partial_{\phi}$ with $G= 2 \chi M e^{a/2}$. Thus we found the first integrals in this case as $I_1 = - E_{\mathcal{L}} = 0, I_2 =  \frac{1}{2} f_0 M^2 e^{(a-b)/2} a'$ and $I_3 = - M^2 e^{(a-b)/2} \phi' - 2 \chi M e^{a/2}$.

\bigskip

\begin{figure}[H]
    \centering
    \subfigure{
        \includegraphics[width=0.3\textwidth]{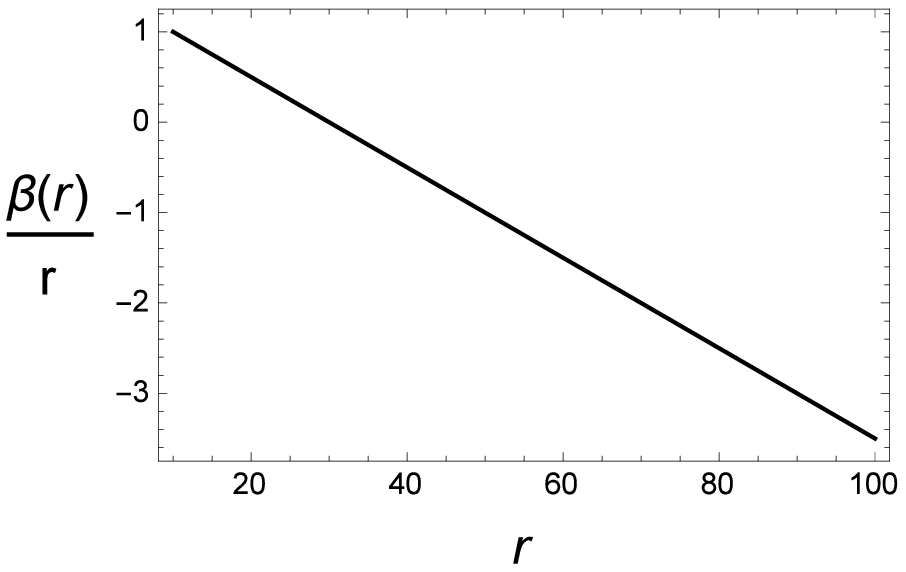}
        \label{fig:subfig5} }
    \subfigure{
        \includegraphics[width=0.3\textwidth]{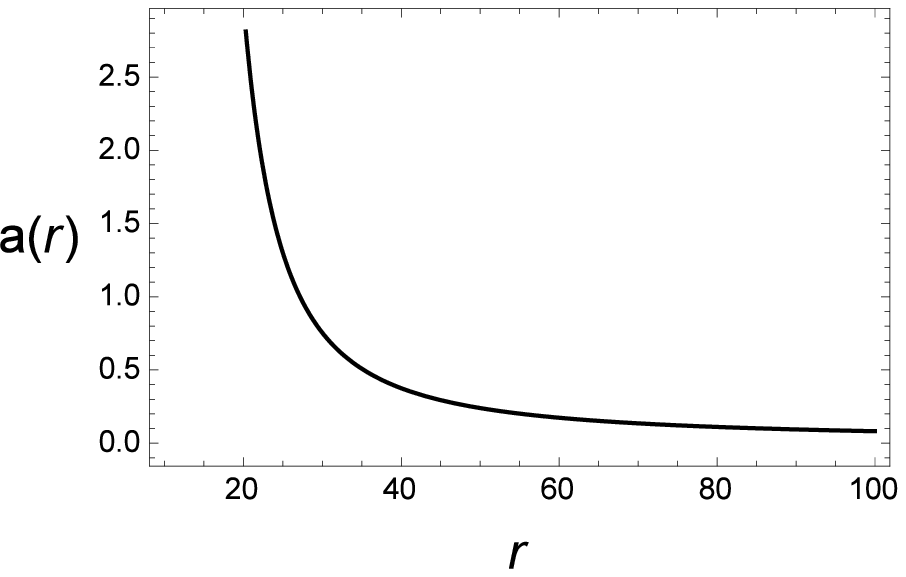}
        \label{fig:subfig6}}
        \subfigure{
            \includegraphics[width=0.3\textwidth]{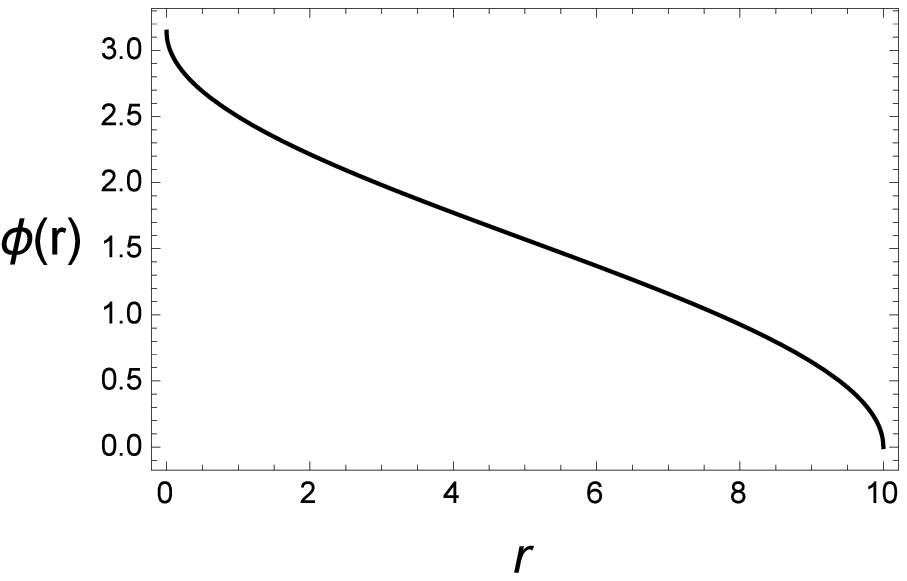}
        \label{fig:subfig7}}
    \caption[Optional caption for list of figures 5-8]{Shape function divided by $r$, redshift function and scalar field versus the radial coordinate for $\ell=1$, $I_{2}=-1.5$, $f_{0}=-1$ and $r_{0}=10$}
    \label{Fig1}
\end{figure}

\begin{figure}[H]
    \centering
    \subfigure{
        \includegraphics[width=0.3\textwidth]{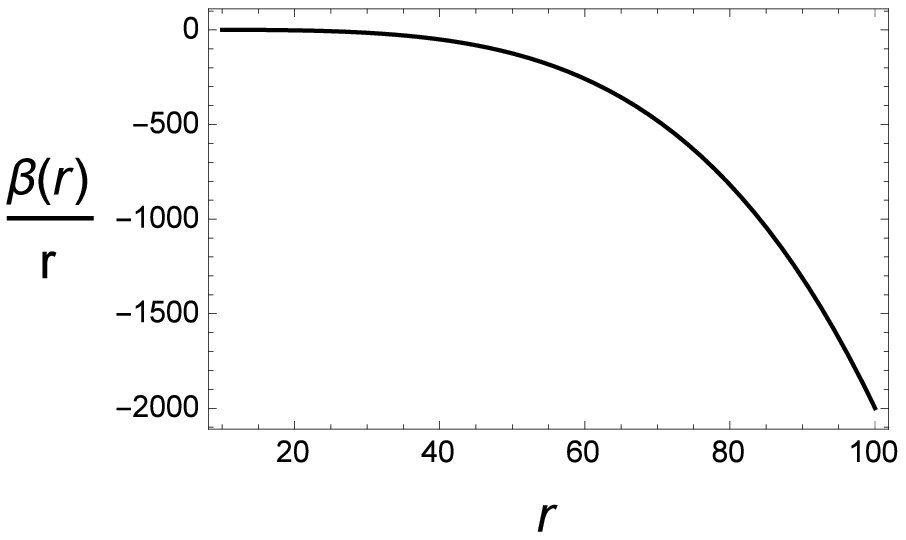}
        \label{fig:subfig11}    }
    \subfigure{
        \includegraphics[width=0.3\textwidth]{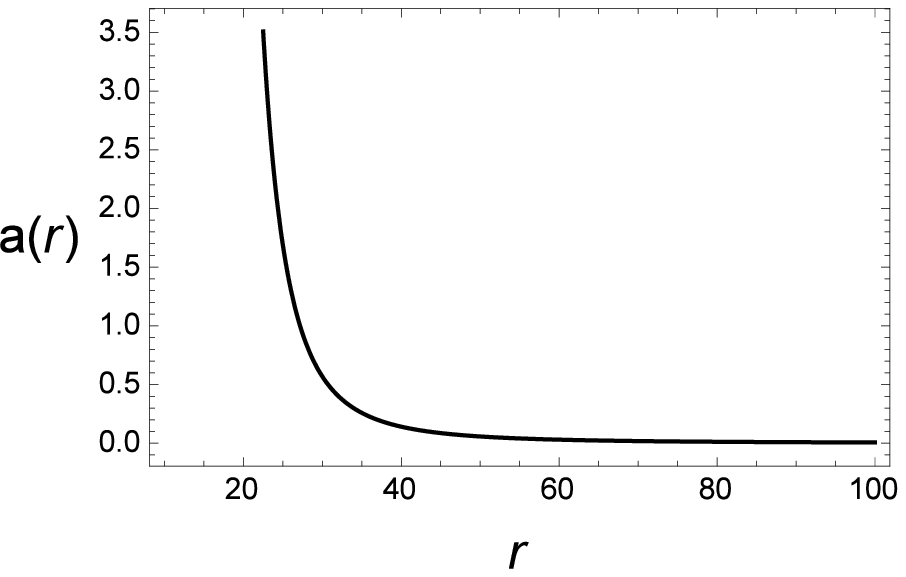}
        \label{fig:subfig12}}
    \subfigure{
        \includegraphics[width=0.3\textwidth]{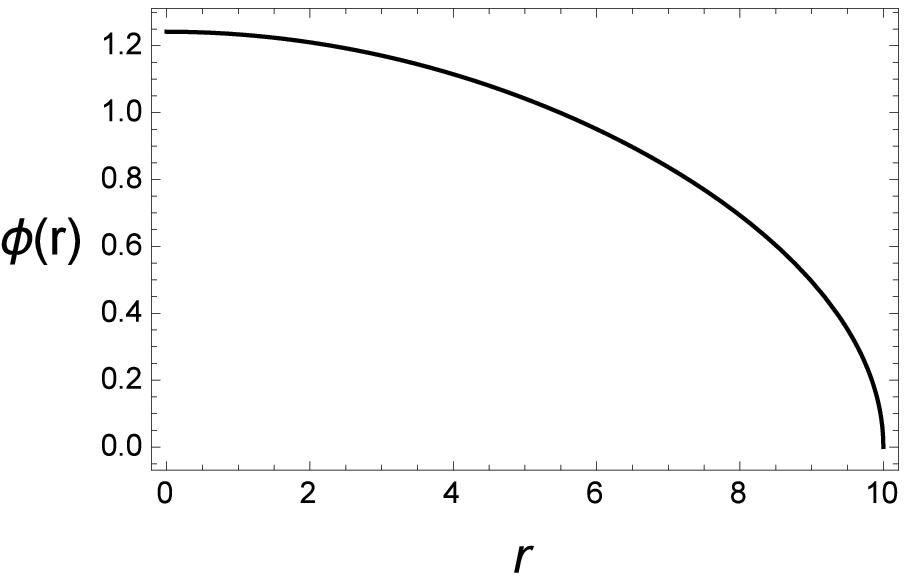}
        \label{fig:subfig13}}
    \caption[Optional caption for list of figures 5-8]{Shape function divided by $r$, redshift function and scalar field versus the radial coordinate for $\ell=1$, $I_{2}=-3$, $f_{0}=-1$  and $r_{0}=10$}
    \label{Fig2}
\end{figure}

{\bf (i.c)}. When $g(\phi) = - f(\phi)$ and $f(\phi) = f_0$ (const.), the components of Noether symmetry generator ${\bf X}$ are
\begin{eqnarray}
& & \xi = c_1 + c_2 r, \quad \eta^1 = 2 c_2 \left( 1 + \ln M \right) - 2 c_3 \phi - 2 c_4, \quad \eta^2 = -2 c_2 \ln M + 2 c_3 \phi + 2 c_4, \nonumber \\& & \eta^3 = M \left( -c_2 \ln M + c_3 \phi + c_4 \right), \quad \eta^4 = c_3 f_0 a + c_5, \quad G = c_6. \label{ngsc-2}
\end{eqnarray}
Then the Noether symmetries are ${\bf X}_1, {\bf X}_2$ given in (\ref{ngsv-1}) and
\begin{eqnarray} \label{ngsv-2}
& & {\bf X}_3 =  \partial_{\phi}, \qquad {\bf X}_4 = \phi {\bf X}_2 + \frac{1}{2} f_0 a {\bf X}_3, \quad {\bf X}_5 = \frac{r}{2} {\bf X}_1 - (\ln M){\bf X}_2 + \partial_a
\end{eqnarray}
with the non-vanishing Lie brackets
\begin{eqnarray} \label{ngsv-2-Lb}
& & \left[ {\bf X}_1, {\bf X}_5 \right] = \frac{1}{2} {\bf X}_1, \qquad \left[ {\bf X}_2, {\bf X}_4 \right] = - \frac{1}{2} f_0 {\bf X}_3, \qquad \left[ {\bf X}_2, {\bf X}_5 \right] = - \frac{1}{2} {\bf X}_2, \\& & \left[ {\bf X}_3, {\bf X}_4 \right] = {\bf X}_2, \quad \qquad \left[ {\bf X}_4, {\bf X}_5 \right] =- \frac{1}{2}\phi {\bf X}_2 - \frac{1}{2} f_0 \left( 1 + \ln M \right) {\bf X}_3 .
\end{eqnarray}
The corresponding Noether constants are
\begin{eqnarray}
& & I_1 = -  E_{\mathcal{L}}, \quad I_2 =  \frac{f_0}{2}  M^2 e^{(a-b)/2} a', \quad I_3 = - M^2  e^{(a-b)/2} \phi', \label{fint-2-1} \\& & I_4 = I_2 \phi + \frac{1}{2} f_0 I_3 a, \qquad I_5 = - I_2 (\ln M) + f_0 e^{(a-b)/2} M M',  \label{fint-2-2}
\end{eqnarray}
where $I_1 =0$ because of the fact $E_{\mathcal{L}} = 0$, which yields
\begin{eqnarray}
& & e^{b(r)} = M M' a' + M'^2 - \frac{M^2}{2 f_0} \phi'^2 . \label{eb-f0}
\end{eqnarray}
Using the first integral $I_2 =  \frac{f_0}{2}  M^2 e^{(a-b)/2} a'$, we have
\begin{eqnarray}
& & e^{(b-a)/2} = \frac{f_0 M^2}{I_2} \frac{u'}{u},
\end{eqnarray}
where $I_2 \neq 0$ and $u \equiv e^{a/2}$. Then, putting the above term into the the first integral $I_5 = - I_2 (\ln M) + f_0 e^{(a-b)/2} M M'$, it becomes
\begin{eqnarray}
& & \frac{u'}{u} = \frac{I_2 \frac{M'}{M}}{I_2 \ln M + I_5},
\end{eqnarray}
which has the solution $\ln u = I_2 \ln M + I_5$, that is, $e^{a(r)} = e^{2 I_5} M^{2 I_2}$ or $a(r) = 2 ( I_2 \ln M + I_5)$. Taking $I_5 = - I_2 \ln M_0$,  one can write that
\begin{eqnarray}
& & e^{a(r)} = \left( \frac{M}{M_0}\right)^{2 I_2},
\end{eqnarray}
where $M(r_0) = M_0$. Thus, together with the first integral $I_3 = - M^2  e^{(a-b)/2} \phi'$, the Eq. (\ref{eb-f0}) yields
\begin{eqnarray}
& & e^{b(r)} = \frac{( 2 I_2 + 1)  M'^2}{1 - \ell (M/M_0)^{-2(I_2 +1)}},
\end{eqnarray}
where $\ell = -I_3^2 / (2 f_0 M_0^2)$. The shape function of the wormhole becomes
\begin{eqnarray}
& & \beta (r) = r \left( 1- \frac{ \left[ 1 - \ell (M/M_0)^{-2(I_2 +1)} \right]}{ ( 2 I_2 + 1)  M'^2 } \right),
\end{eqnarray}
where $f_0 \neq 0, M' \neq 0$ and $I_2 \neq -1/2$. The function $M(r)$ is independent of $\theta$ and $\phi$ because of isotropy, and it must have the limit $M(r) \sim r$ as $r \rightarrow 0$. The form of the function $M(r)$ can be determined by use of the geodesic deviation equation (see page 50 of Ref.~\cite{Ellis}), which gives that the solutions for $M(r)$ with the appropriate limit behaviour $M(r) \sim r$ as $r \rightarrow 0$ are $M(r) = \epsilon^{-1} \sin(\epsilon r)$ if $k=1$, $r$ if $k=0$ and $\epsilon^{-1} \sinh (\epsilon r)$ if $k=-1$, where $k$ is the curvature parameter and $\epsilon$ is a dimensional constant. In the limit case $M(r)= r$, the metric coefficients, the shape function and the scalar field take the form
\begin{eqnarray}
& & e^{a(r)} = \left( \frac{r}{r_0} \right)^{2 I_2}, \qquad  e^{b(r)} = \frac{( 2 I_2 + 1) } {1 - \ell \left( \frac{r}{r_0} \right)^{-2(I_2 +1)}}, \label{w1-a}\\ & & \beta (r) = r \left[ 1- \frac{1}{ ( 2 I_2 + 1)} + \frac{ \ell}{ ( 2 I_2 + 1)} \left( \frac{r}{r_0} \right)^{-2(I_2 +1)} \right], \label{w1-b}  \\& & \phi(r) = - \frac{1}{(I_2 + 1)} \sqrt{\frac{2I_2 + 1}{2 f_0}} \arctan \left[ \sqrt{\frac{1}{\ell} \left( \frac{r}{r_0} \right)^{2(I_2 +1)} -1 } \right]. \label{w1-phi}
\end{eqnarray}
The metric coefficients (\ref{w1-a}) and the shape function (\ref{w1-b}) are very similar to the wormhole solution found by Cataldo et. al \cite{cataldo} under the assumption of isotropic pressure. It can be noted that in this solution we must set $\ell=1$ to satisfy the wormhole condition at the throat $\beta(r=r_{0})=r_{0}$. Figs.~\ref{Fig1} and \ref{Fig2} show the behaviour of this solution for $\beta(r)/r$, $a(r)$ and $\phi(r)$ for different values of the parameters $I_{2}$,  $r_{0}$ and $f_{0}$. In order to have a real and positive scalar field, the parameters must satisfy $I_2 <-1$ and $f_{0}<0$. Moreover, the scalar field must lie outside the wormhole in the region $[0,r_{0}]$, otherwise it will become negative or complex. We can see from the pictures that the wormhole obeys the asymptotic flatness in both cases since as $r\rightarrow \infty$, a de Sitter space-time is recovered.

%\begin{figure}[H]
%   \centering
%   \subfigure{
%       \includegraphics[width=0.3\textwidth]{shapefI2_15r0_1}
%       \label{fig:subfig8} }
%   \subfigure{
%       \includegraphics[width=0.3\textwidth]{resdshiftI2_1_5r010}
%       \label{fig:subfig9}}
%   \subfigure{
%       \includegraphics[width=0.3\textwidth]{scalarfieldI2_1_5r010}
%       \label{fig:subfig10}}
%   \caption[Optional caption for list of figures 5-8]{General Caption of subfigures 5-8}
%   \label{fig:subfigureExample3}
%\end{figure}

For $M(r) = \epsilon^{-1} \sin(\epsilon r)$, we find that
\begin{eqnarray}
& & e^{a(r)} = \left( \frac{\sin(\epsilon r)}{\sin(\epsilon r_0)} \right)^{2 I_2}, \qquad  e^{b(r)} = \frac{( 2 I_2 + 1) \cos^2 (\epsilon r) } {1 - \ell \left( \frac{\sin(\epsilon r)}{\sin(\epsilon r_0)} \right)^{-2(I_2 +1)}}, \label{w2-a}\\ & & \beta (r) = r \left[ 1- \frac{1}{ ( 2 I_2 + 1) \cos^2 (\epsilon r) } \left\{ 1- \left( \ell \frac{\sin(\epsilon r_0)}{\sin(\epsilon r)} \right)^{2(I_2 +1)} \right\} \right], \label{w2-b}
\end{eqnarray}
which is a new wormhole solution as far as we know. One can find a similar solution for $M(r) = \epsilon^{-1} \sinh (\epsilon r)$.\\
If $g(\phi) = 0$, which corresponds to the sub cases (i.a.1) and (i.b.1), then the Noether symmetries are ${\bf X}_1$ and ${\bf X}_2$ given in (\ref{ngsv-1}) for arbitrary $f(\phi)$.

{\bf (i.d)}. If $g(\phi) = 0$ and $f(\phi)=(f_0 + f_1 \phi )^2$, where $f_1 \neq 0$, the solution of Noether symmetry equations (\ref{ngs-eqs}) are
\begin{eqnarray}
& & \xi = c_1, \quad \eta^1 = -2 c_2 - 2 c_3 \left( 1 + \frac{1}{M} +  \ln M \right), \quad \eta^2 = 2 c_2 + 2 c_3 \left( \frac{1}{M} +  \ln M \right), \nonumber \\& & \eta^3 = c_2 M + c_3 \left( M \ln M + 1 \right), \quad \eta^4 = \frac{c_3}{2 f_1} ( f_0 + f_1 \phi), \quad G = c_4, \label{ngsc-3}
\end{eqnarray}
that is, the ${\bf X}_1, {\bf X}_2$ by (\ref{ngsv-1}) and
\begin{eqnarray}
& & {\bf X}_3 = \left( \ln M + \frac{1}{M} \right) {\bf X}_2  - \partial_a + \frac{1}{4 f_1} \left( f_0 + f_1 \phi \right) \partial_{\phi},
\end{eqnarray}
are Noether symmetries. Here there is only non-vanishing Lie bracket due to the ${\bf X}_1, {\bf X}_2$ and ${\bf X}_3$ as
\begin{eqnarray} \label{ngsc-3-Lb}
& & \left[ {\bf X}_2, {\bf X}_3 \right] = \frac{1}{2}\left( 1- \frac{1}{M} \right) {\bf X}_2.
\end{eqnarray}
Here the first integrals for ${\bf X}_1, {\bf X}_2$ and ${\bf X}_3$ are $I_1 = - E_{\mathcal{L}} = 0, I_2 =  \frac{1}{2} (f_0 + f_1 \phi^2) M^2 e^{(a-b)/2} a'$ and
\begin{equation}
I_3 = I_2 \left( \frac{1}{M} + \ln M \right) - (f_0 +  f_1 \phi)^2 \left( e^{(a-b)/2} M M' -e^{a/2} M  \right) -\frac{1}{4 f_1} (f_0 + f_1 \phi) M^2 e^{(a-b)/2} \phi' .
\end{equation}

If $g(\phi) = 0$ and $f(\phi) = f_0 = const$,  which corresponds to the sub case (i.a.4) or (i.b.4), then the components of Noether symmetry generator ${\bf X}$ are same as (\ref{ngsc-2}), but the gauge function $G$ is obtained by $G = -2 c_2 f_0 M e^{a/2} + c_6$. Therefore in this case the Noether constant $I_5$ related with ${\bf X}_5$ has the form
\begin{eqnarray}
& & I_5 = - I_2 (\ln M) + f_0 e^{(a-b)/2} M M' + f_0 M e^{a/2}.
\end{eqnarray}

{\bf Case (ii):} $V(\phi) = V_0$ (constant potential).

When $g(\phi) = - f(\phi)$ and $f(\phi) = f_0$ or $g(\phi) = 0$ and $f(\phi) = f_0$ (constant), we have only two Noether symmetries ${\bf X}_1 = \partial_r$ and ${\bf X}_2 = \partial_{\phi}$ which gives rise to the Noether constants $I_ 1 = - E_{\mathcal{L}} \equiv 0$ and $I_2 = - M^2 e^{(a-b)/2} \phi'$. Thus, the relation $E_{\mathcal{L}}=0$ gives
\begin{eqnarray}
& & e^{b(r)} = \frac{ e^a ( M M' a' + M'^2 ) }{ e^a \left( 1- \frac{V_0 M^2}{f_0} \right) + \frac{I_2^2}{2 f_0 M^2} }.
\end{eqnarray}
For $M(r) = r$, it follows that
\begin{eqnarray}
& & e^{b(r)} = \frac{ e^a ( r a' + 1 ) }{ e^a \left( 1- \frac{V_0 r^2}{f_0} \right) + \frac{I_2^2}{2 f_0 r^2} },
\end{eqnarray}
which yields the shape function of the wormhole as
\begin{eqnarray}
& & \beta (r) = \frac{ r \left[ e^a \left( r a' + \frac{V_0 r^2}{f_0} \right) - \frac{I_2^2}{2 f_0 r^2} \right]}{e^a (r a' +1)}. \label{shape-2}
\end{eqnarray}
Considering the specific redshift function given by $a(r) = 0$ \cite{Bohmer:2011si}, and inserting this condition into the Eq.~(\ref{shape-2}) the shape function reduces to the form
\begin{eqnarray}
& & \beta (r) = \frac{r}{f_0}  \left( V_0 r^2 - \frac{I_2^2}{2 r^2} \right). \label{shape-3}
\end{eqnarray}
Now, we take into account the Noether first integral $I_2 = - r^2 e^{(a-b)/2} \phi'$, and find that
\begin{eqnarray}
& & \phi (r) = I_2 \sqrt{\frac{f_0}{2}} \ln \left[ 2 \left( \sqrt{\frac{2}{f_0}} \frac{I_2}{r} \sqrt{ 1- \frac{V_0 r^2}{f_0} + \frac{I_2^2}{2 f_0 r^2}} + \frac{I_2^2}{f_0 r^2} +1 \right) \right] + \phi_0, \label{phi-1}
\end{eqnarray}
where $\phi_0$ is an integration constant.\\
For $V(\phi) = V_0 \phi^n$, where $n= 1,2,...$, there is only one Noether symmetry ${\bf X}_1 = \partial_r$, which is trivial solution and corresponds to the first integral $I = - E_{\mathcal{L}} = 0$.

\section{Discussion and Conclusions}

Teleparallel theories and their extensions have gained a lot of 
attention in recent years as alternative gravitation frameworks.
Additionally, in literature, these theories have been studied by
their couplings, both minimal and non-minimal, with torsion. Our
choice of action generalizes and extends most of the earlier models
of teleparallel gravity by assuming an additional coupling between the scalar field and a boundary term $B$ which is related to the divergence of torsion. Moreover, this new coupling allows to formulate  a theory that, under some constraints, allows to construct quintessence, teleparallel dark energy and non-minimally coupled scalar field to the Ricci scalar theories. The latter has been used in the literature to study spherically symmetric configurations as wormholes (see Ref.~\cite{Barcelo:2000zf}) obtaining the existence of traversable wormholes supported by non-minimally coupled scalars.

Here we derived new wormhole
solutions according to the  Morris and
Thorne paradigm. In this connection, we adopt the Noether Symmetry
Approach solving the governing equations  to obtain
symmetry generators, gauge function and the metric coefficients of
the wormhole geometry. To perform this analysis, we focused on two forms
of potential function, i.e. $V=0$ and $V=constant$. In the former
case, we obtained five symmetry generators while in latter case, only
two. For the vanishing potential, we explicitly found two different wormhole solutions in the case $f(\phi)=-g(\phi)=constant$, i.e., within a quintessence model.  The first wormhole solution is similar to that found by Cataldo et. al \cite{cataldo} under the assumption of isotropic pressure, whereas the second wormhole solution is, in our knowledge, a new wormhole solution. 

It can be seen that these solutions respect the asymptotic flatness as $r\rightarrow \infty$. In  the case where a constant potential is assumed, we also found an explicit wormhole solution for the quitessence case ($g=0$ and $f=constant$). For a power-law form of potential, only one generator exists which corresponds to a trivial solution.

It might be interesting to perform a similar analysis by assuming a more general action as, for example, to  include also coupling between the scalar field and the teleparallel torsion Gauss Bonnet term $T_{G}$ or a Gauss Bonnet boundary term $B_{G}$ (see for example Ref.~\cite{Bahamonde:2016kba}). These will be the arguments of future work. 

\section*{Acknowledgments}
S.C. wishes to acknowledge the support by  Istituto 
Nazionale di Fisica Nucleare, Sezione di Napoli, Italy, iniziative 
specifiche TEONGRAV and QGSKY. S. B. and S. C.  also acknowledge 
the COST Action CA15117 (CANTATA), supported by COST (European 
Cooperation in Science and Technology). S.B. is supported by the Comisión Nacional de Investigación Científica y Tecnológica (Becas Chile Grant No. 72150066).

\end{document}